\documentstyle[aps,multicol]{revtex}
\renewcommand{\narrowtext}{\begin{multicols}{2} \global\columnwidth20.5pc}
\renewcommand{\widetext}{\end{multicols} \global\columnwidth42.5pc}
\begin{document}

\title{Comparing conductance quantization in quantum
wires and Quantum Hall systems
}

\author{Anton Yu. Alekseev$^{* \dagger \ddagger}$,
 Vadim V. Cheianov$^{* \dagger \ddagger }$, 
J\"{u}rg Fr\"{o}hlich$^{\dagger}$}

\address{$^*$
Insitute of Theoretical Physics, Uppsala University,
Box 803, S-75108, Uppsala, Sweden}

\address{ $^\dagger$
Insitut f\"{u}r Theoretische Physik, ETH-H\"{o}nggerberg,
CH-8093, Z\"{u}rich, Switzerland}

\address{$^\ddagger $
Steklov Mathematical Institute, Fontanka 27, 191011,
St.Petersburg, Russia}

\date{July 1996}
\maketitle

{\tightenlines
\begin{abstract}
We propose a new calculation of the DC
conductance of a 1-dimensional
electron system described by the Luttinger model.
Our approach is based on the ideas of Landauer and B\"{u}ttiker
and on the methods of current algebra.
We analyse in detail the way in which the system can be 
coupled to external reservoirs.
This determines whether the conductance is renormalized or not.
We show that although a quantum wire and a Fractional Quantum Hall
system are described by the same effective theory, their coupling
to external reservoirs is different. 
As a consequence, the conductance in the
wire is quantized in integer units of $e^2/h$ per spin
orientation  whereas
the Hall conductance allows for fractional quantization.
\end{abstract}
}


\narrowtext

Recent experiments on transport properties of
quantum wires \cite{vW,THS} attracted  new interest
to the problem of  conductance  of a 1-dimensional
electron gas.

Since the work of Apel and Rice \cite{AR}, where the authors
computed the conductance in the 1-dimensional Luttinger
model,  it was 
believed that the effects of interaction in the one-channel
quantum wire should lower the conductance with respect to
the value 
\begin{equation} \label{cond}
\sigma = 2 \frac{e^2}{h}
\end{equation}
predicted by
the Landauer-B\"{u}ttiker formula \cite{But} for the case
of non-interacting electrons. However,
experimental data suggest that this renormalization of
conductance does not actually take place \cite{THS}.

At the moment there exist a number of theoretical arguments
\cite{Pon,MS,SS,Kaw,Kaw2} intended
to explain the non-renormalization of the conductance in 1-dimensional
electron gas. One of them \cite{Pon,MS} is based on the idea
that the conductance of the 1-channel quantum wire is completely determined
by the structure of the leads to which the wire is attached. So far
as the electrons in the leads form a Landau Fermi-liquid 
one chooses to model the leads  by two semi-infinite
1-dimensional 1-channel non-interacting Fermi liquids
whereas the wire is modeled by
a Luttinger Fermi-liquid. Due to strong non-locality of
current-current correlators in 1-dimensional systems the calculation
based on the Kubo formula leads to the nonrenormalized result (\ref{cond}).

Another approach \cite{Kaw,Kaw2} is based on the observation that
the dielectric constant in the Luttinger Fermi-liquid increases
with increasing  strength of interaction.
This fact changes the definition of the voltage drop which results in
the cancellation
of the renormalization factor in the conductance. In this approach one
does not consider the reservoirs to which the wire is attached and
does not explain how the voltage drop is measured in experiment.

Both approaches  mentioned above are based on  linear response theory
and its main tool, the Kubo formula, in their calculation of conductance.
Yet there is another, in many ways  more intuitive  approach
based on ideas of Landauer and B\"{u}ttiker \cite{But}. In this paper we show 
how to apply these ideas to a  Luttinger Fermi-liquid type
channel.

We study how the Luttinger liquid interacts with external
reservoirs. In order to clarify our analysis of  quantum wires,
we provide a parallel treatment of the Fractional Quantum Hall system
which is described by the same effective Hamiltonian
\cite{Wen,Fr}. It appears that
the electron transfer between the Luttinger liquid and external
reservoirs in these two systems is organized in a different way.
This difference  leads to the integer quantization of conductances in
quantum wires in contrast to the fractional quantization of
Hall conductances.
   
We replace  appealing but clearly unrealistic (1-dimensional) 
models of the leads \cite{Pon,MS} by a more universal consideration
which relies only on the fact that the leads interact with the
Luttinger liquid via electron transfer.
It is difficult to
compare our results to  \cite{Kaw,Kaw2} as they do not
consider the interaction to reservoirs at all and in our approach
this is the central issue which determines whether the conductance is
renormalized or not.

It is well known that a 1-dimensional interacting electron
system  is effectively described by the Luttinger model \cite{Hal}.
For simplicity we consider spinless fermions. This makes
the comparison to the spin polarized Quantum Hall system
more straightforward. Polarization of fermions can be
easily recovered in the final expression for conductance.

The bosonized Lagrangian density  of the Luttinger model looks
as follows \cite{Hal}:

\begin{equation} \label{LL}
{\cal L}=\frac{hv_F}{4}\left(\frac{1}{v_F^2}
(\partial_t \phi)^2 -
(1+g)(\partial_x \phi)^2\right),
\end{equation}
where $v_F$ is the Fermi velocity,  
$n=\partial_x \phi$ is the particle density,
$I=e \partial_t \phi$ is the electric current, $g$ is an effective coupling
constant. The corresponding Hamiltonian  is given by 
\begin{equation} \label{HL}
H_0=\frac{hv_F}{4}\int dx :\left(\frac{1}{v_F^2} (\partial_t \phi)^2 +
(1+g)(\partial_x \phi)^2\right):\ ,\end{equation}
where $::$ stands for the standard Wick ordering.
 The model with $g=0$
describes a non-interacting 1~-~dimensional Fermi-liquid.

The Hamiltonian (\ref{HL}) can be expressed in terms
of left-moving and right-moving
currents:

\begin{equation}
H_0=\frac{h}{2v_F}\int dx :(I_L^2 +I_R^2): \ ,
\end{equation}
where

\begin{eqnarray}
I_L=\frac{1}{2}(\partial_t \phi + v_F \sqrt{1+g}\ \partial_x \phi), 
\nonumber \\
I_R=-\frac{1}{2}(\partial_t \phi - v_F \sqrt{1+g}\ \partial_x \phi),
\end{eqnarray}
so that

\begin{equation}
I=e (I_L-I_R) \ , \ n=\frac{1}{v_F\sqrt{1+g}}(I_L+I_R).
\end{equation}

Let us remark that the same effective Hamiltonian describes
edge excitations of a spin polarized incompressible
 Quantum Hall fluid on a cylinder
\cite{Wen} with one chiral channel of edge excitations
per boundary.
Then $eI_L$ and $eI_R$ are the edge currents on the two boundaries
of the cylinder. The filling factor of the Quantum Hall fluid
is related to the effective coupling
of the Luttinger model via:

\begin{equation} \label{rel}
\nu=\frac{1}{\sqrt{1+g}}.
\end{equation}
Incompressibility requires that $\nu^{-1}$ is an odd integer.

In order to measure conductances in the Luttinger model we should
couple it to two reservoirs with different chemical potentials \cite{But}.
This can be done in two different ways. We shall see that one
way is realized in the Quantum Hall system. The corresponding
conductance should be interpreted as the Hall conductance
\begin{equation} \label{*}
\sigma_H=\nu \ \frac{e^2}{h}.
\end{equation} 
The other way of coupling the Luttinger model to reservoirs
is realized in thin wires and leads to the non-renormalized value
\begin{equation} \label{**}
\sigma= \frac{e^2}{h}.
\end{equation}

The coupling  of reservoirs to the Luttinger liquid can only be organized
via electron transfer. We therefore must identify excitations in
the Luttinger liquid which correspond to electrons
and assign to them chemical
potentials of reservoirs according to the conditions of the
measurement. 

This can be done naturally for incompressible Quantum Hall fluids.
In this case the chiral  channels are spatially separated by the
bulk of the sample and the reservoirs are attached directly to each
of them. Mathematically this is expressed by adding an extra term to
the Hamiltonian:

\begin{equation} \label{Hmu}
H_{\rm Hall}=H_0+ \mu_L N_L +\mu_R N_R,
\end{equation}
where $\mu_L$ and $\mu_R$ are the chemical potentials of the 
reservoirs $eV=\mu_R-\mu_L$, $N_L$ and $N_R$ are 
conserved electron  numbers
in the left-moving and right-moving channels
(corresponding to the  two different edges of the sample): 

\begin{eqnarray}
N_L= \int dx \ n_L \ , \ n_L= \frac{1}{v_F} \nu \ I_L; \nonumber \\
N_R= \int dx \ n_R \ , \ n_R= \frac{1}{v_F} \nu \ I_R.
\end{eqnarray}
The expectation value of the operator (\ref{Hmu}) can be interpreted
as a free energy of the system at zero temperature.
We minimize  the energy functional which yields the uniform ground
state current
\begin{equation} \label{calcul}
 I=e\ \nu \left( \frac{\mu_R}{h} - \frac{\mu_L}{h}\right)=
 \nu \ \frac{e^2}{h}V. \end{equation}
Thus, the Hall conductance is given by formula (\ref{*}).
This result would correspond to the renormalized conductance
$$\sigma= \frac{1}{\sqrt{1+g}}\frac{e^2}{h}$$
in the Luttinger model.

In the Quantum Hall system the left- and right-moving edge excitations
 of charge $e$ are represented by operators 
\begin{eqnarray}
\Psi_L(x)=:\exp \{ i\pi(\sqrt{1+g} \phi(x) + 
\frac{1}{v_F}\int^x \partial_t \phi dx) \}:\ , \nonumber \\
\Psi_R(x)=:\exp \{ i\pi(\sqrt{1+g} \phi(x) - 
\frac{1}{v_F}\int^x \partial_t \phi dx) \}: \ .
\end{eqnarray}
These operators correspond to particles with Fermi statistics only when
$\sqrt{1+g}$ is an odd integer \cite{Wen,Fr}.
These values of the filling factor
\begin{equation}
\nu=\frac{1}{2m+1}
\end{equation}
correspond to Laughlin's plateaux \cite{L} in the theory of Fractional Quantum
Hall Effect.

Next we turn to the analysis of quantum wires.
For generic values of the coupling constant $g$ neither left- nor
right-moving  edge excitations can be identified with electrons.
Hence, we can not assign particular values of chemical potentials
to these excitation branches. Yet, there always exist other
fermionic excitations of charge $e$ that describe the physical electrons
in the Luttinger model. They
are created by applying the operators $\Psi_+$, $\Psi_-$ given by
\begin{eqnarray}
\Psi_+(x)=:\exp \{ i\pi( \phi(x) + 
\frac{1}{v_F}\int^x \partial_t \phi dx) \}: \ , \nonumber \\
\Psi_-(x)=:\exp \{ i\pi( \phi(x) - 
\frac{1}{v_F}\int^x \partial_t \phi dx) \}: \ .
\end{eqnarray}
These operators coincide with $\Psi_L$, $\Psi_R$ in the non-interacting
model ($g=0$). The particle densities corresponding to the operators
$\Psi_+$ and $\Psi_-$ are given by
\begin{eqnarray}
n_+=\frac{1+\sqrt{1+g}}{2} n_L + \frac{1-\sqrt{1+g}}{2} n_R, \nonumber \\
n_-=\frac{1-\sqrt{1+g}}{2} n_L + \frac{1+\sqrt{1+g}}{2} n_R.
\end{eqnarray}

It is natural to view the chemical potentials of the reservoirs 
as the variables conjugate to the
conserved charges
\begin{equation}
N_+=\int dx \ n_+ \ , \ N_-=\int dx \ n_-
\end{equation}
which measure the total number of electrons created by powers of
$\Psi_+$ and $\Psi_-$,
respectively.
If  the quantum channel in the wire
is ballistic and adiabatic, the chemical potential of the left reservoir  
is conjugate to $N_+$ whereas the chemical potential of the
 right reservoir is conjugate to $N_-$. The counterpart of the
Hamiltonian (\ref{Hmu}) therefore is given by
\begin{equation}
H_{\rm Wire}=H_0+ \mu_L N_+ +\mu_R N_-.
\end{equation}
A calculation
similar to the one in (\ref{calcul}) shows that the ground state
of $H_{\rm Wire}$ carries a current
\begin{equation}
I= e \left( \frac{\mu_R}{h}-\frac{\mu_L}{h}\right) =\frac{e^2}{h} V
\end{equation}
corresponding to (\ref{**}).
Two spin polarizations of electrons can be included in the final answer
for conductance of  a single quantum channel by adding a factor of $2$
which leads to (\ref{cond}).

We conclude that although the Quantum Hall system and the quantum wire
are described by the same effective Hamiltonian, the way in which
they are coupled to reservoirs in  transport measurements is different.
Mathematically, this is reflected in using chemical potentials
$\mu_L$($\mu_R$) of left(right) reservoirs conjugate variables to
 different conserved charges $N_L$($N_R$), for
Quantum Hall fluids, and $N_+$($N_-$) for quantum wires.
As a consequence, in the Quantum Hall system the conductance depends on the 
effective coupling constant related to the filling factor by (\ref{rel}),
but this is not the case for the quantum wire.

\widetext

\end{document}